\documentclass[%
 preprint,
 amsmath,amssymb,
 aps,
onecolumn
]{revtex4-2}
\usepackage{textcomp, gensymb}
\usepackage{physics}
\usepackage{color}
\usepackage{bm}
\usepackage{amsmath}
\usepackage{amsfonts}
\usepackage{booktabs}
\usepackage{rotating}
\usepackage{soul}
\usepackage{siunitx}

\newcommand{\ra}[1]{\renewcommand{\arraystretch}{#1}}
\begin{document}

\preprint{APS/123-QED}

\title{Simulation of a Three-Nucleons System Transition on Quantum Circuits}

\author{Luca Nigro}
\author{Carlo Barbieri}
\author{Enrico Prati}%
 \email{enrico.prati@unimi.it}
\affiliation{%
Dipartimento di Fisica ``Aldo Pontremoli",\\ Università degli studi di Milano, via Celoria 16, I-20133 Milan, Italy
}

\date{\today}

\begin{abstract}
\noindent Quantum computers have proven to be effective in simulating many quantum systems. Simulating nuclear processes and state preparation poses significant challenges, even for traditional supercomputers. This study demonstrates the feasibility of a complete simulation of a nuclear transition, including the preparation of both ground and first excited states. To tackle the complexity of strong interactions between two and three nucleons, the states are modeled on the tritium nucleus. Both the initial and final states are represented using quantum circuits with variational quantum algorithms and inductive biases. Describing the spin-isospin states requires four qubits, and a parameterized quantum circuit that exploits a total of 16 parameters is initialized. The estimated energy has a relative error of approximately $2\%$ for the ground state and about $10\%$ for the first excited state of the system. The simulation estimates the transition probability between the two states as a function of the dipole polarization angle. 
This work marks a first step towards leveraging digital quantum computers to simulate nuclear physics.
\end{abstract}

\maketitle
\section{Introduction}
\noindent Quantum simulation is an emerging field that aims at studying 
quantum systems
with quantum hardware to reach a level of accuracy unattainable with classical computers. Quantum computers \cite{gill2022quantum, henriet2020quantum, huang2020superconducting, bruzewicz2019trapped, manzalini2020topological, ferraro2020all, de2023silicon} are proving increasingly effective to provide such hardware~\cite{Gilbert19}. 
One of the most developed areas where quantum simulation proves effective involves few-body and many-body search of ground states and dynamics. 
Molecular quantum dynamics provides a paramount example in this respect \cite{ollitrault2021molecular}. Simulation and discovery of materials is a rapidly evolving field that can exploit near-term quantum hardware \cite{clinton2024towards}. 
Recently, quantum algorithms have also been applied to nuclear physics in order to determine nuclear structures and simulate elementary processes~\cite{GarciaRamos, Roggero19, Roggero20C, Roggero20D}. \\
Some of us have already exploited variational and iterative quantum algorithms \cite{agliardi, maronese,corli} and the search of ground states \cite{CRS4} by different approaches based on adiabatic quantum computing \cite{rocutto,moro,noe}.
Variational Quantum Eigensolvers (VQE)~\cite{Peruzzo14} made it possible to build states of light nuclei using suitably tuned quantum circuits~\cite{cerezo, hobday22, hobday22bis}, with the aim of computing ground state energies.
On the other hand, quantum algorithms enable to compute transition probabilities between states encoded by such quantum circuits. Explicitly, the computation is carried by another circuit consisting in turn of a unitary operator. The observable responsible of such transition is not necessarily unitary, even if it can be expressed as the sum of unitary terms like Pauli operators that are not closed under addition. Therefore, one needs to embed it into a unitary operator by some method so to be processed by the quantum computer. Quantum algorithms such as the linear combination of unitaries (LCU) can solve the issue of restoring unitarity~\cite{Childs, Roggero19, Roggero20C, Roggero20D} and permit the implementation of the circuit.\\
Transition probabilities are fundamental also for understanding phenomena involving atomic nuclei. For example dipole polarizabilities are intimately linked to nuclear radii and the properties of nucleonic matter, eventually affecting the merging of compact astrophysical objects~\cite{Hagen16WkCa,Lattimer2000}. More recently, it has been pointed out how a detailed description of the strong many-body correlations in nuclei is necessary to explain electroweak processes such has $\beta$ decay~\cite{Gysbers2019Bdec} or the neutrinoless $\beta\beta$ decay that is employed for searching physics beyond the standard model~\cite{Gomez2023VoBBdec}.  
In spite of important advances in modelling the interaction of neutrinos with nuclei~\cite{Roggero19,Roggero20D}, there is currently no implementation on a quantum computer of the whole pipeline of a nuclear transition process: that is, involving the preparation of both the initial and final nuclear states as well as the transition mechanism. 
Here, we consider a simplified model of the nucleus of tritium in which the proton and the two neutrons are fixed in space, so to demonstrate the full quantum computing pipeline for simulating a nuclear reaction. We employ a pionless effective field theory~($\pi\!\!\!/$EFT)~\cite{eft} Hamiltonian that provides a controllable low-momentum expansion of the nuclear force, consistent with the symmetries of quantum chromodynamcis~(QCD).
The spatial localization preserves the complex spin-isospin structure of the strong nuclear force while reducing the necessary qubits to a number suitable for a proof-of-principle investigation. At the same time, discrete excited states are produced--even if they are experimentally absent for three-nucleons systems--that we use to simulate a M1 transition to the ground state.
We demonstrate that a quantum computer is capable of simulating an entire nuclear problem by addressing all the building blocks involved in the pipeline.
We apply a VQE algorithm that allows to determine the ground state by minimizing the energy function. Next, we build an excited state from a similar minimization problem by involving an additional cost function that accounts for the orthogonality between eigenstates.
Computed the matrix elements of an excitation operator -- operation which in turn can be carried employing the quantum resources themselves, the quantum computer is enabled to calculate transition probabilities since coefficients required for linear combination of unitaries (LCU) method are now embedded in the quantum algorithm.
The LCU method is then used to evaluate the transition probability as a function of the tilt angle.
We conclude that quantum algorithms are sufficiently mature to perform simulations in nuclear physics.


\section{Methods}
\noindent The system under consideration is inspired by the tritium nucleus (triton) and the process involves the transition between two eigenstates. To demonstrate the method, we arbitrarily choose the ground state and the first excited state of such a three-nucleons system.
\begin{figure*}[tbp]
    \centering
    \includegraphics[width=0.8\textwidth]{ 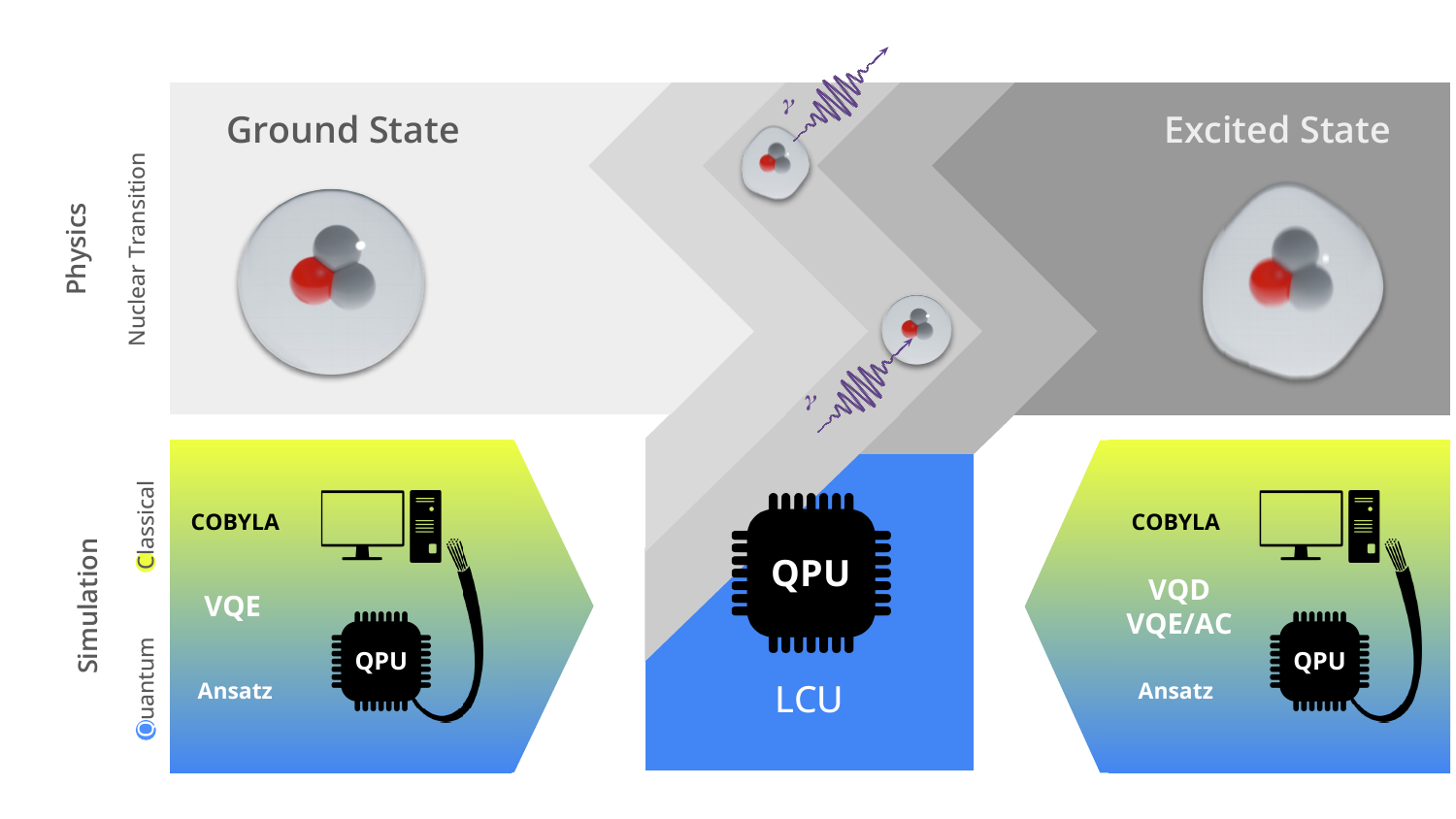}\hspace{50pt}

    \caption{Overview of the physical phenomenon under investigation (top) and the computational pipeline to simulate it on a quantum computer (bottom). The process consists of the excitation of the nucleus of tritium (triton) from the ground state to the first excited state. $\widehat{O}$ is the operator responsible for such a transition. By using variational quantum algorithms, both the ground state of the triton (on the left), and its excited state are implemented. The former requires the variational quantum eigensolver (VQE), the latter the variational quantum deflation (VQD) and the variational quantum eigensolver with automatically-adjusted constraints (VQE/AC).} After such quantum states are prepared, one implements the excitation operator on a quantum circuit with the help of the linear combination of unitaries (LCU) method. The blue areas indicate the use of a quantum computer (or quantum processing unit, QPU), and the faded green to blue areas represent a hybrid quantum-classical computation (the optimizer, COBYLA, performs classical computation).
    \label{fig:general_scheme}
\end{figure*}
 In Figure~\ref{fig:general_scheme} it is sketched the rationale of the approach. The upper part represents aspects of the ingredients of the nuclear physics process, consisting of the excitation (or equivalently, its de-excitation) between two eigenstates, while the lower part represents the corresponding computational strategy of the quantum computer. The state preparation of qubits encodes the relevant information of the nuclear states under consideration. Next, it is simulated the action of the interaction responsible for the transition between such quantum states, evaluating the transition probability.\\
In the following, we introduce the nuclear model of the ground state of the triton and, afterwards, the preparation of a quantum register encoding the three-nucleons eigenstates. Next, in the Section Results, we exploit such formalization to perform the quantum simulation and obtain the transition probability with a gate-model quantum computer. Since real hardware has not been used, no error mitigation technique has been applied.

\subsection{Tritium nuclear model}
To establish the feasibility of simulating nuclear physics processes on a quantum computer, we use the $\pi\!\!\!/$EFT~\cite{eft} on a lattice~\cite{lattice_eft} which is suitable for demonstrating the generality of the approach~\cite{Roggero20D}.
Each nucleon has a spin-isospin state. By using second quantization formalism, we are able to encode the possible states accounting for the whole nucleus with only four qubits.
In order to do so, we consider three nucleons on a $2\times 2$ lattice 
with periodic boundary conditions. One nucleon is fixed on a specific lattice size, hence we need two qubits for each remaining nucleon.
If the static nucleon is placed on lattice site 1, the Hamiltonian of the tritium nucleus is expressed by
\begin{equation}
\begin{split}
    H &= -t\sum_{f=1}^{2}\sum_{\langle i,j \rangle} c_{i,f}^\dag c_{j,f} + 2dtA + U\sum_{i=1}\sum_{f<f'}^{2}n_{i,f}n_{i,f'}\\ &+V\sum_{f<f'<f''}^{2}\sum_{i=1}n_{i,f}n_{i,f'}n_{i,f''} +U\sum_{f=1}^{2}n_{1,f} + V\sum_{f<f'}^{2}n_{1,f}n_{1,f'}
\end{split}.
\end{equation}
The case of $V=-4U$, together with a Jordan-Wigner transformation let us write the same Hamiltonian as
\begin{equation}\label{eq:hamiltonian}
    H = 8t + \frac{U}{2} - 2t\sum_{k=1}^4 X_k - \frac{U}{4}\left(Z_1Z_4 + Z_2Z_3\right) - \frac{U}{4}\sum_{i< j < k}Z_i Z_j Z_k.
\end{equation}
For a more in-depth discussion, one should refer to \cite{Roggero20D}.
\noindent In the following, we use the numerical values of $t=1\,\si{\mega\electronvolt}$ and $U=-7\,\si{\mega\electronvolt}$
that reproduce the actual $\pi\!\!\!/$EFT nuclear force for the three nucleons placed at neighboring sites~\cite{Roggero20D}.  The objective is to find the ground state, hence we need to minimize the expectation value of such a Hamiltonian.
In order to test the minimum set of ingredients, we consider excitations to higher eigenstates of the tritium model. More specifically, we focus on the first excited state of the model. This is done by solving a differently constrained problem which takes into account the orthogonality with the ground state previously found. In the next Section, we are going to define such minimization problems in more detail. 

\subsection{Variational quantum eigensolver (VQE)}
\begin{figure*}[tbp]
    \centering
    \includegraphics[width=0.5\textwidth]{ 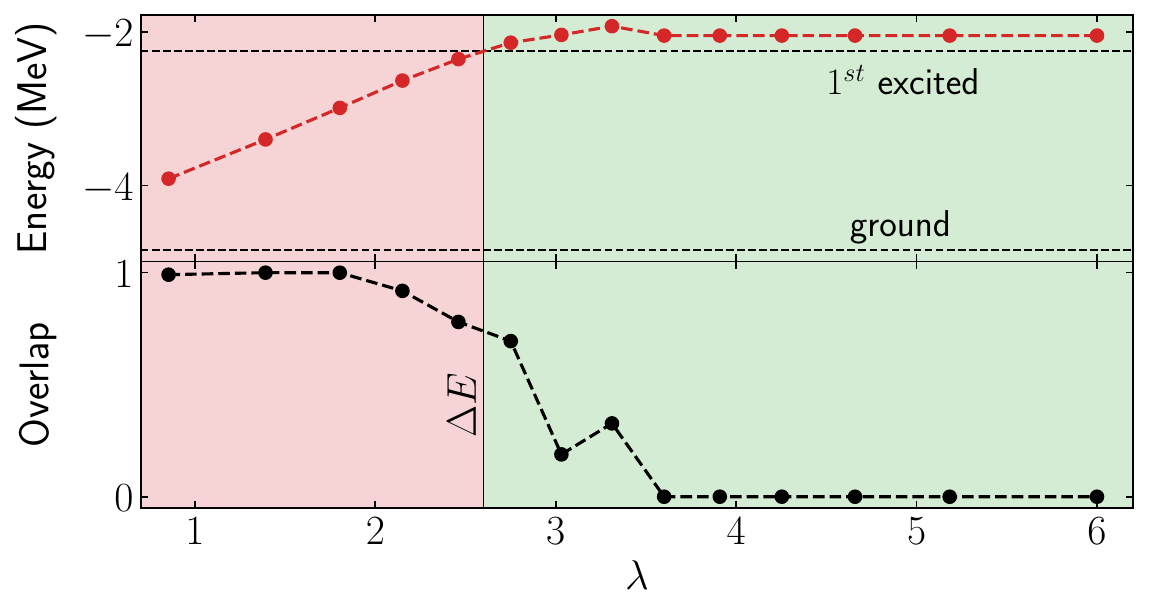}
    \caption{The minimization problem solved by the variational quantum deflation (VQD) algorithm has a penalty term that accounts for the orthogonality of the state with a previously-found ground state. The coefficient $\lambda$ is the weight of such constraint in the minimization problem. The choice of $\lambda$ is self-correcting, meaning that by choosing an incorrect value $\lambda=\gamma - E_0 \le \Delta E$, will cause the algorithm to find a minimum proportional to $\gamma$. From the upper plot, it is evident the linear behaviour of the cost function for $\lambda < \Delta E$ (red shaded area), and then a constant optimal value for $\lambda \ge \Delta E$ (green shaded area). The lower plot keeps track of the overlap of the candidate excited state with the ground state. For the results of this article, $\lambda$ is set to be equal to $\lambda=4\,\si{\mega\electronvolt}$.}
    \label{fig:lambda_tuning}
\end{figure*}

The variational quantum eigensolver~(VQE) is a hybrid variational algorithm consisting of a quantum eigensolver and a classical optimizer.
We consider a parameterized ansatz state $\ket{\psi(\bm{\theta})}$ sufficiently complex to reproduce the fundamental properties of the system to be simulated.
Here $\bm{\theta}$ is a vector of real parameters. The optimal parameters can be found by solving the related variational problem.\\
The ground state is, by definition, the state of minimum energy, hence the minimization problem
\begin{equation}\label{eq:variational_energy}
    \min_{\bm{\theta}} E(\bm{\theta}) \qq{ with } E(\bm{\theta})=\expval{H}{\psi(\bm{\theta})}
\end{equation}
allows us to find the best estimate for the ground state. In addition, by imposing orthogonality with such a ground state, the new minimum energy eigenstate becomes the first excited state.\\
The first method exploited to find the first excited state involves adding a penalty function (known as deflation term) to Equation~\ref{eq:variational_energy}, while the second one delegates such a constraint to the optimizer. The former is known as variational quantum deflation~(VQD)~\cite{higgott19}, while the latter is the variational quantum eigensolver under automatically-adjusted constraints~(VQE/AC)~\cite{Gocho23}.
The minimization problem solved by the VQD can be expressed as
\begin{equation}\label{eq:variational_deflation}
        \min_{\bm{\theta}} L_E(\bm{\theta})\qquad L_E(\bm{\theta}) = E(\bm{\theta}) + \lambda\,\abs{\braket{g}{\psi(\bm{\theta})}}^2
\end{equation}
where $\ket{g}$ is the ground state and $\lambda$ is a tunable hyper-parameter.
If the ansatz is efficiently expressive, it is sufficient to use $\lambda > \Delta E$, where $\Delta E$ is the energy gap $\Delta E = E_1 - E_0$ between the first two energy levels. 
However, the choice of $\lambda$ is self-correcting, as choosing an incorrect $\lambda = \gamma - E_0 \le \Delta E$ will cause the algorithm to find a minimum $L_E(\bm{\theta})=\gamma$~\cite{higgott19}. Such a behaviour is reported in Figure~\ref{fig:lambda_tuning}, where $\lambda=4\,\si{\mega\electronvolt}$ appears to be a good choice.

\begin{figure*}[tbp]
    \centering
    \includegraphics[width=0.85\textwidth]{ 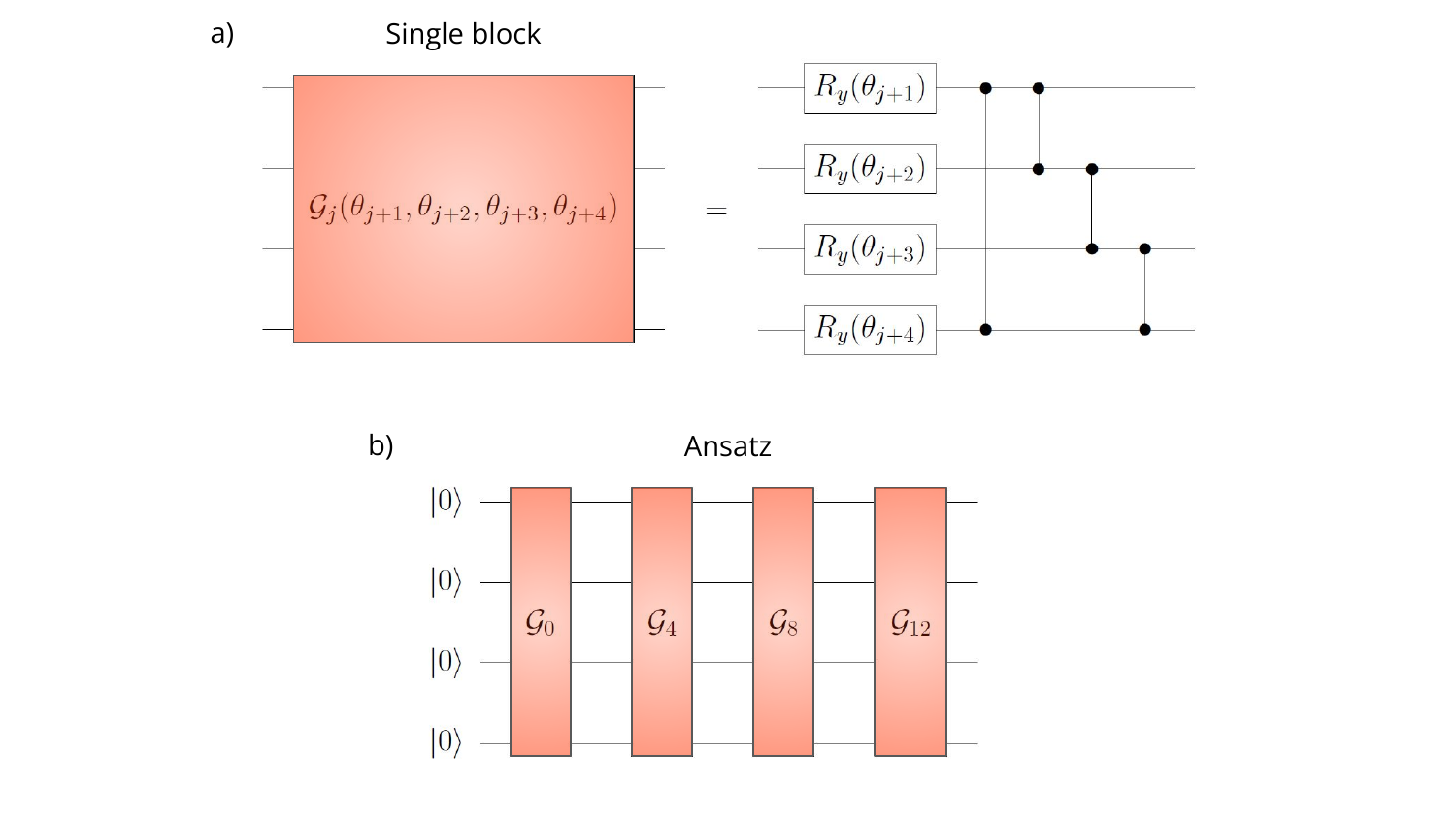}
    \caption{Parameterized quantum circuit that reproduces the quantum state of the tritium nucleus. It is made of four qubits. {\bf a)} The single block $\mathcal{G}_j$ is made of a rotation layer, where each angle becomes a parameter, and a circular entanglement scheme, made of controlled-$Z$ gates. {\bf b)} The ansatz is made of four repetitions of the single block, i.e. $\mathcal{G}_{12}\mathcal{G}_8\mathcal{G}_4\mathcal{G}_0\ket{0000}$. It requires each qubit to be initialized in the state $\ket{0}$ and it has a total of $16$ parameters, since each block $\mathcal{G}_j$ has 4 parameters. }\label{fig:ansatz}
\end{figure*}

On the contrary, the VQE/AC method does not require such calibration since it handles the constraint of orthogonality on the optimizer directly. Hence the minimization problem is again the one shown in Equation~\ref{eq:variational_energy}.
Typically, an ansatz is formed by repeating blocks, each one made of two layers, namely one with local rotations and the other with entangling gates. In our work, the rotation layers consist of single-qubit $y$-rotations acting on each qubit and the entangling layer accounts for circular entanglement, significantly extending previous attempts involving only one rotation angle per layer, with only two blocks. As discussed later, by doubling the number of blocks and by adding more parameters, a more accurate ground state energy is found. 
Furthermore, a deeper ansatz allows us to use the same circuit for both the ground state and the first excited state while keeping the number of trainable parameters as low as possible.
The choice of the number of blocks is arbitrary, so the best compromise between accuracy and depth, involving more computation, is found. The best compromise consists of using four blocks, as it is the least amount that significantly improves the results (see Supplementary Figure~S1).
If the ansatz is made of only one block the energy landscape has many local minima from which it is hard to escape, even with the imposition of the orthogonality constraint in the VQD and {VQE/AC} methods. Therefore as shown in Supplementary Figure~S1, only a sufficient depth of the ansatz secures that the eigenstates are found starting from the ground state to higher-energy states.
For insufficient depth energy levels may be found in reversed order when the true ground state is missed by the VQE.
The single block definition and the final ansatz are shown in Figure~\ref{fig:ansatz}.

\subsection{Simulation of the transition}
Now, we turn the attention to the quantum operator that triggers the transition between the ground state and the excited state. For clarity, we call it the excitation operator, but the reader should notice that since it is real-valued, hence symmetric, the inverse process is also described by the same operator. In order to implement the action of such operator onto the quantum register, we exploit the linear combination of unitaries method, that allows to implement any operator that can be expressed as a sum of unitaries.
\noindent Following Ref.~\cite{Roggero20C}, the most general excitation operator in first quantization is
\begin{equation}\label{eq:excitation_operator}
    \widehat{O} = \alpha\mathbb{I} +\beta X + \gamma Z
\end{equation}
where $\alpha$, $\beta$ and $\gamma$ are real coefficients.
As mentioned in Additional Section A available as supplementary material, the observable responsible for the transition is the magnetic dipole moment $\vb{m}\cdot\bm{\xi}$. In order to calculate the values of the coefficients $\alpha$, $\beta$ and $\gamma$, we need to evaluate the expectation values in Equation~{SE7}. There are two distinct methods that may be employed: a classical approach leveraging a simulator backend, and a quantum approach that utilizes a quantum computer to directly estimate these values.
When using a simulator backend, the operator $\vb{m}\cdot\bm{\xi}$ is expressed as a matrix, the quantum states as state vectors, and the expectation value 
$\bra{g} \vb{m}\cdot\bm{\xi}\ket{g}$ is evaluated classically via matrix multiplications.\\
Alternatively, the expectation value can be evaluated on a quantum computer. To achieve this, 
think of the ansatz as a unitary operator $U_{\bm\theta}$. By substituting the parameter vector $\bm\theta$ with the optimal values for the ground state ($\bm\theta_g$) and excited state ($\bm\theta_e$) found with the variational algorithms, we can write each expectation value as
$\bra{0} U^\dag_{\bm\theta_g}( \vb{m}\cdot\bm{\xi}) U_{\bm\theta_g}\ket{0}$.
Now decompose the operator $U^\dag_{\bm\theta_g}( \vb{m}\cdot\bm{\xi}) U_{\bm\theta_g}$ into Pauli operators, estimate them singularly and then sum everything classically (the expectation value of a linear combination of operators is the linear combination of the expectation values of such operators).
Since this last approach requires to execute a quantum circuit multiple times and collect measurement outcomes, we preferred the classical approach.
In both cases, $\theta_g$ and $\theta_e$ play a key role and they are determined in the first part of this work.\\
In second quantization, it can be expressed as
\begin{equation}\label{eq:tilde_excitation_operator}
    \widetilde{O} = \left(\alpha+\gamma\right) c_0^\dag c_0 + \left(\alpha-\gamma\right) c_1^\dag c_1 + \beta \left(c_0^\dag c_1 + c_1^\dag c_0\right).
\end{equation}
If we represent the creation/annihilation operators in terms of Pauli operator using the Jordan-Wigner transformation, the excitation operator becomes 
\begin{equation}
    \widetilde{O} = \alpha\mathbb{I} + \gamma Z_0 + \frac{\beta}{2}\left(X_0X_1 + Y_0Y_1\right).
\end{equation}
Restricted to the relevant subspace, $\widetilde{O}$ is equivalent to
\begin{equation}\label{eq:triton_excitation_operator}
    \overline{O} = \alpha\mathbb{I} + \frac{\gamma}{2} (Z_0 - Z_1) + \frac{\beta}{2}\left(X_0X_1 + Y_0Y_1\right).
\end{equation}
Such operator consist of a sum of Pauli, hence unitary operators, but $\overline{O}$ itself is not necessarily unitary.

\subsection{The linear combination of unitaries (LCU) algorithm}\label{sec::lcu}
The algorithm that implements such an operator is expressed by a quantum circuit as follows. 
The evolution generated by the operator $\overline{O}$ can be simulated in a non-deterministic way using the linear combination of unitaries (LCU) method~\cite{Childs}. The only requirement is that it can be expressed as a (finite) sum of unitaries:
\begin{equation}
    O = \sum_{j=0}^L \mu_j U_j
\end{equation}
The coefficients $\mu_j$s are assumed to be positive without loss of generality since a phase can be subsumed into the respective unitary operators $U_j$.\\
 \begin{figure*}[tbp]
    \centering
    \includegraphics[width=\textwidth]{ 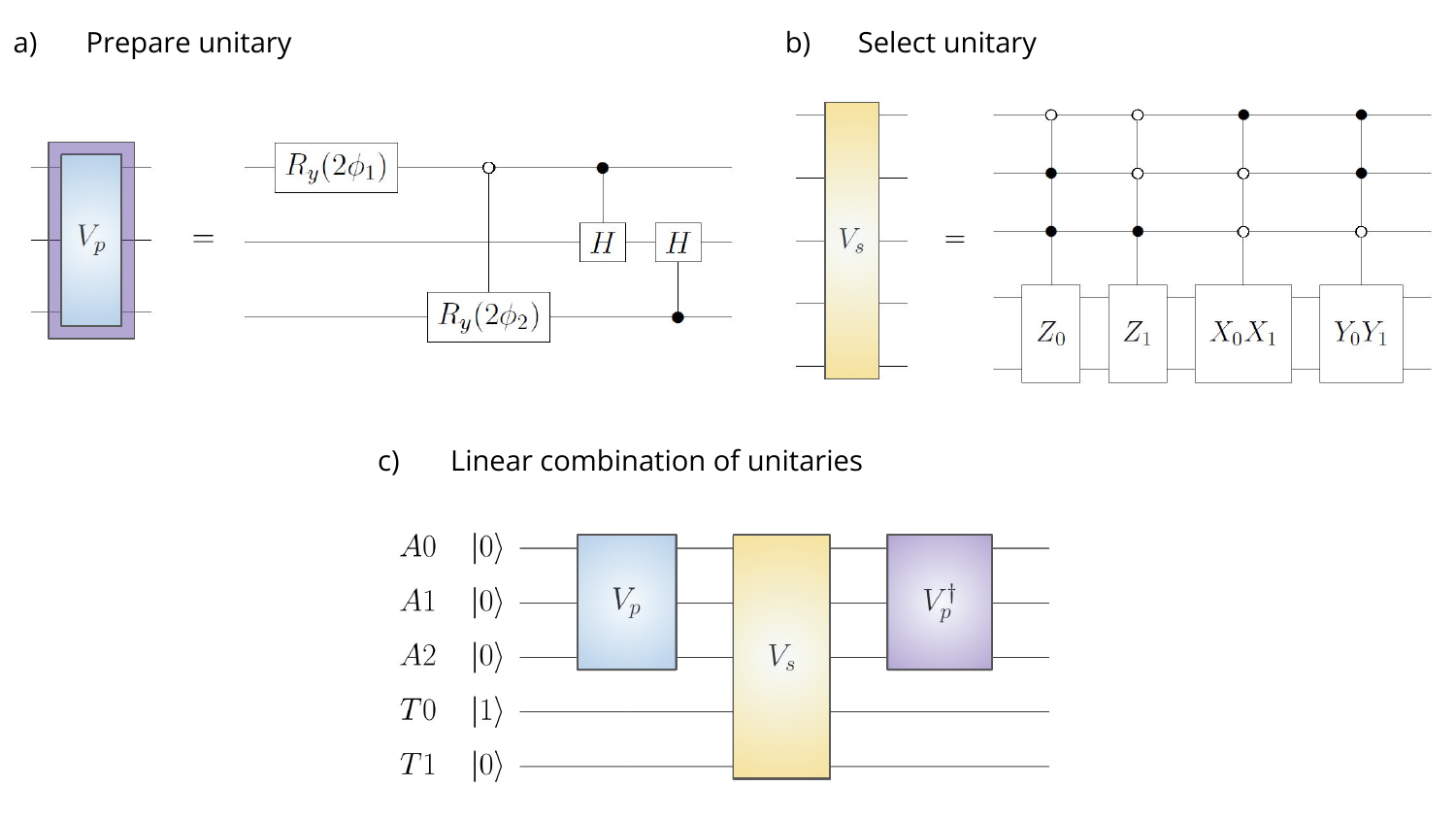}
    \caption{Quantum gates used in the linear combination of unitaries (LCU) method. It requires the definition of a prepare unitary acting on the ancilla register and a select unitary acting on the target register. {\bf a)} Implementation of the prepare unitary $V_p$ (in blue) for the excitation operator in Equation~\ref{eq:triton_excitation_operator} acting on three ancilla qubits. The color violet indicates the dagger version of the unitary operator $V_p$. {\bf b)} Implementation of the select unitary $V_s$ (in yellow) for the same excitation operator acting on the ancilla register and the two target qubits encoding the state of the triton. {\bf c)} Linear combination of unitaries (LCU) circuit diagram. The ancilla qubits are $A0$, $A1$ and $A2$, whereas the target qubits are named $T0$ and $T1$.
    The ancilla register is initialized in the state $\ket{000}$ and the method is considered successful if and only if such ancilla register at the end is measured to be again in the state $\ket{000}$. The target register is initialized in the state $\ket{10}$ encoding the ground state in second quantization. With such formalism, we want to study the probability of having the target qubits in the state $\ket{01}$, and we call it transition probability $P_t$.}\label{fig:lcu_circuit}
\end{figure*}
It is useful for later purposes to define the 1-norm of the coefficient vector
\begin{equation}\label{eq:spectral_norm}
\Lambda := \sum_{j=0}^L \mu_j .
\end{equation}
For example, for the excitation operator above, we have $\Lambda = \abs{\alpha} + \abs{\beta} + \abs{\gamma}$.
The algorithm requires the implementation of a \textit{prepare} unitary and a \textit{select} unitary. The former is defined as~\cite{Childs, Roggero20C}
\begin{equation}\label{eq:prepare_unitary}
    V_p\ket{0} = \sum_{j=0}^L \sqrt{\frac{\mu_j}{\Lambda}}\ket{j} ,
\end{equation}
whereas the latter
\begin{equation}\label{eq:select_unitary}
    V_s = \sum_{j=0}^L\ketbra{j}\otimes U_j.
\end{equation}
The required prepare unitary is shown in Figure~\ref{fig:lcu_circuit}(a), with a rotation $R_y(2\phi_1)$, where 
\begin{equation}\label{eq:phi1}
\phi_1 = \arcsin(\sqrt{\frac{\abs{\beta}}{\Lambda}}),
\end{equation}
acting on the first qubit. Contrary to the deuteron case \cite{Roggero20C}, the second gate applied is a zero-controlled rotation $R_y(2\phi_2)$, where
\begin{equation}\label{eq:phi2}
\phi_2 = \arcsin(\sqrt{\frac{\abs{\gamma}}{\abs{\alpha}+\abs{\gamma}}}).
\end{equation}
As a matter of fact, the excitation operator for the deuterium nucleus has a simpler form of Equation~\ref{eq:excitation_operator}. Despite having three distinct coefficients, it holds that
\begin{equation}
    \gamma = -\alpha\quad\Rightarrow \quad \phi_2 = \arcsin(1/\sqrt{2}) = \pi/4
\end{equation}
hence the rotation $R_y(\pi/2)$ acts as a Hadamard gate. Instead, here we generalize such transition to the case where $\alpha,\,\beta,\,\gamma$ and their moduli are distinct. The circuit implementation of the select unitary follows directly from Equation~\ref{eq:select_unitary}.
Its circuit implementation is shown in Figure~\ref{fig:lcu_circuit}(b).\\
The number of ancilla qubits needed for the LCU scheme is determined by the number of unitaries involved in the sum of Equation~\ref{eq:excitation_operator}. As from Equation~\ref{eq:triton_excitation_operator}, the operator is composed of a sum of $5$ unitaries, hence the ancilla register has $M = \lceil\log(5)\rceil=3$ qubits. Such condition ensures to perform a conditional operation for each and every one of them.
The number of terms in Equation~\ref{eq:tilde_excitation_operator} scale at worst as $\mathcal{O}(N^2)$, where $N$ is the number of dynamical nucleons. After the Jordan-Wigner transformation, in Equation~\ref{eq:triton_excitation_operator}, the number of terms $L$ scale again with $\mathcal{O}(N^2)$. Although the number of ancilla qubits required by the LCU algorithm is $M = \lceil\log(L)\rceil$, which means $M = \mathcal{O}(2\log(N))$, the number of needed operations scales as $\mathcal{O}(N^2)$ in the worst case scenario. Hence there is no obvious quantum advantage.
Having addressed the encoding of the problem, we now move to the explicit implementation of the transition on a gate-based quantum computer architecture.

\section{Results}
In this Section, we start by reporting on the results of the variational quantum algorithms, which return the ground state and the first excited state of the triton nucleus, respectively. 
The second step consists therefore to use such information to determine the probability of having a transition dictated by the excitation operator $\overline{O}$ of Equation~\ref{eq:excitation_operator}. The results of the simulation of the nuclear transition described by the excitation operator $\overline{O}$ are described in the second Subsection.

\subsection{Determination of the initial and final quantum states with VQE}
\begin{figure}[tbp]
    \centering
    \includegraphics[width=0.45\textwidth]{ 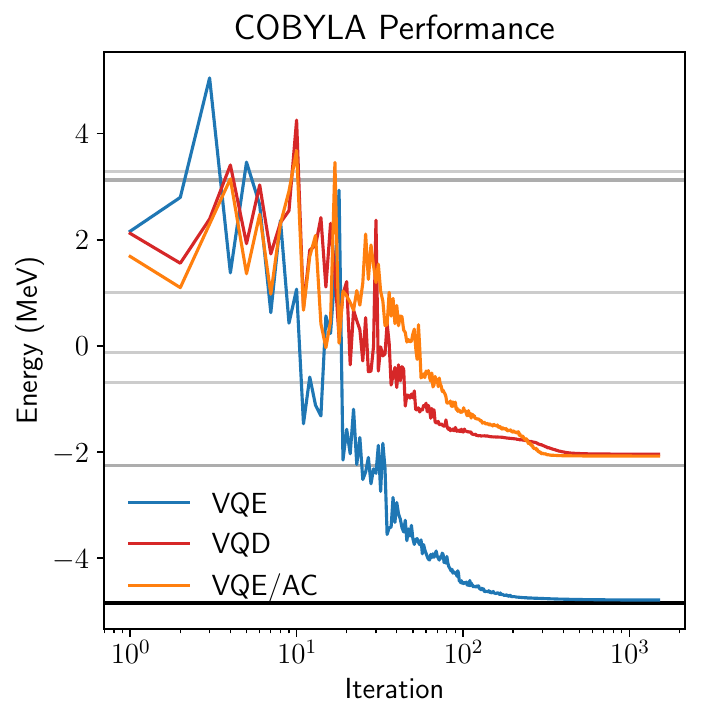}
\caption{Performance of the minimization of the triton energy $\min_{\bm{\theta}} E(\bm{\theta})$ by using three different variational algorithms: vanilla VQE (in blue), VQD (in red), and VQE/AC (in orange).
The VQE performs the minimization of energy in search of the ground state of the triton. The VQD and VQE/AC, on the other hand, minimize the energy with an orthogonality constraint with respect to the ground state previously found. The lowest energy levels of the triton Hamiltonian obtained numerically are shown in gray. The ground state energy is shown in black.}
\label{fig:minimization_vqd_vqeac}
\end{figure}

\noindent The optimizer is the constrained optimization by linear approximation (COBYLA)~\cite{cobyla}, with a maximum of 1500 iterations. The circuits are implemented by Qiskit~\cite{qiskit} through the QASM simulator within the Qiskit Aer framework. The ansatz circuit requires four qubits as from Figure~\ref{fig:ansatz}(b). The ground state is determined by the VQE algorithm after 400 iterations. Instead, the excited states is determined after 300 iterations with VQE/AC and after 450 iterations with VQD, respectively, without achieving the same accuracy.
Figure~\ref{fig:minimization_vqd_vqeac} illustrates the performance of COBYLA over $1500$ iterations. Each colored line represents a different algorithm, superimposed on the energy spectrum  calculated analytically  (grey lines), with the black line indicating the ground energy level. 

\begin{figure*}[tbp]
    \centering
    \includegraphics[width=\textwidth]{ 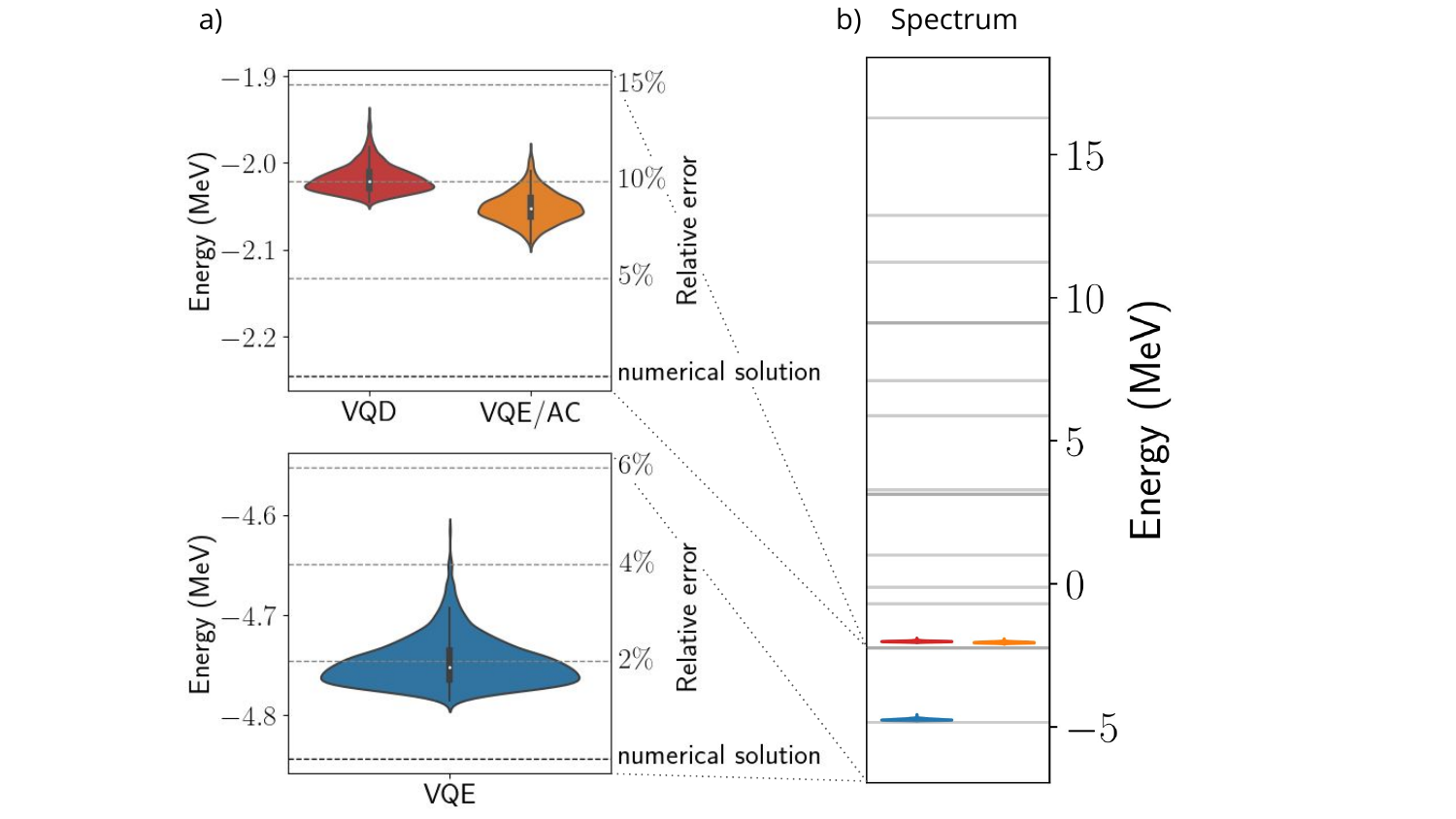}
    \caption{{\bf a)} Estimated expectation values of the Hamiltonian, calculated on the first excited state (top) and on the ground state (bottom). The violin plot refers to the distribution obtained via Monte Carlo simulation sampling the parameters of the ansatz from a normal distribution centered in the best estimate and with standard deviation equal to the error. {\bf b)} Numerical (and classical) energy levels of the triton Hamiltonian together with the quantum estimation.}
    \label{fig:res_violin}
\end{figure*}

The optimal parameters are listed in Table~\ref{tab:optimal_pars}. By assigning such values, we evaluate the expectation value of the Hamiltonian in Equation~\ref{eq:hamiltonian}. To understand how the error in the optimal-parameters estimation propagates to the energy estimation, we adopt a Monte Carlo method, as follows.
We sample a normal distribution centered on the optimal value with precision up to the third decimal place, with a standard deviation of $\sigma = 0.001$. For each sample we then evaluate the energy expectation value. This allows us to retrieve the energy distribution for each variational algorithm used, providing insight into the propagation of error from optimal parameters to energy estimation. The resulting violin plots are reported in Figure~\ref{fig:res_violin}.
The optimal parameters are associated with a relative error in the energy value of $\epsilon_0^\text{VQE} = 1\% $ for the ground state, $\epsilon_1^\text{VQD} = 9\%$ and $\epsilon_1^\text{VQE/AC} = 8\%$ for the first excited state.

\subsection{Transition simulated with the LCU method}
We now turn to the excitation operator $\overline{O}$ which drives the nuclear transition under investigation.
In order to proceed, the coefficients $\alpha, \beta, \gamma$ are determined by using the optimal parameters found above, as described in Additional Section A, available as supplementary material accompanying the online article. 
The excited state chosen for this estimate is that obtained by the VQE/AC algorithm, since it better approximates the energy eigenvalue.
The resulting nuclear reaction is of type M1, where the magnetic dipole moment, characterized by the polarization angle $\vartheta$, is responsible of the transition.  
First, we report on the the  polarization of the magnetic dipole moment lying in the $xz$-plane. Next, the simulation explores three-dimensional space where the polarization direction is described by two angles ($\vartheta, \varphi$).
To evaluate the transition probability at a given polarization angle $\vartheta$ of the magnetic dipole gamma emission, we employ the LCU scheme on a quantum computer. Such algorithm is non-deterministic, succeeding only if the ancilla register is in the state $\ket{0}$ at the end of the computation \cite{Childs}. If not, the computation has to be repeated because such result has to be rejected. The success probability $P_s$ is defined as the ratio of successful runs to total trials. Upon each successful run, a measurement is taken on the target register, which has been initialized in the ground state. If found in the excited state at the end, the simulation indicates a transition. The transition probability $P_t$ is the ratio of runs ending in the excited state to the number of successful runs.\\
The physical operator that governs the transition corresponds to the magnetic dipole moment, which in turn can be parametrized by its polarization angle $\vartheta\in \left[0,\pi\right]$. Accordingly, the coefficients $\alpha(\vartheta), \beta(\vartheta)$ and $\gamma(\vartheta)$ of Equation~\ref{eq:excitation_operator} are functions of the same angle.
Both of probabilities, $P_s$ and $P_t$, are evaluated at some dipole polarization angles $\vartheta$.
\begin{figure*}[tbp]
    \centering
    \includegraphics[width=0.8\textwidth]{ 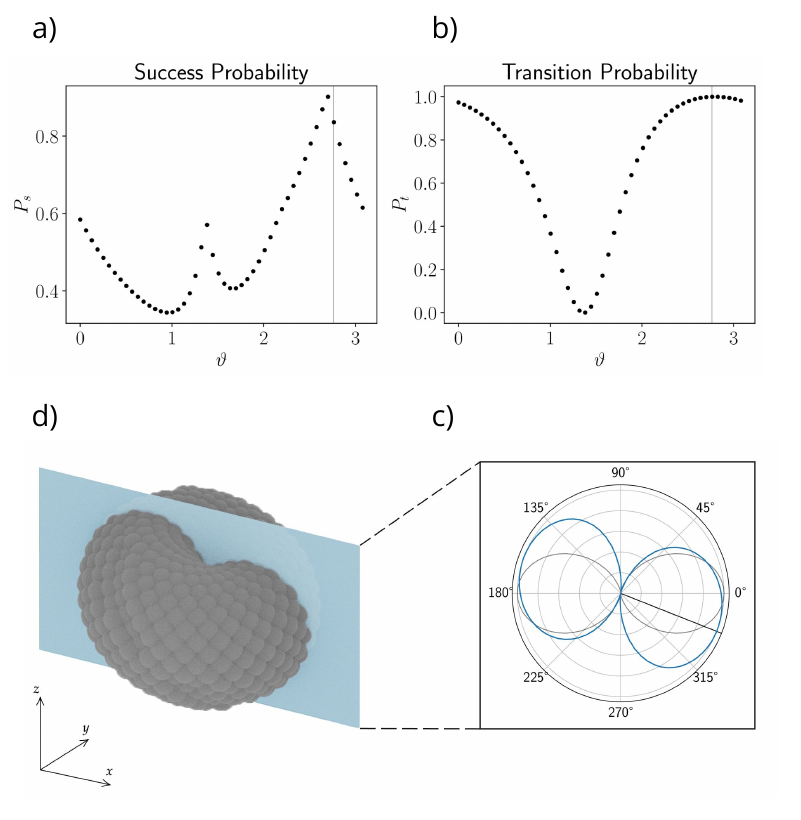}
    \caption{Simulation of the quantum circuit over the QASM simulator provided by Qiskit.
{\bf a)} Success probability of the LCU method applied to the excitation operator responsible for the transition of the triton from the first excited (VQD) state into 
 the ground state. 
 {\bf b)} the transition probability computed by the LCU.
 Both are studied for variable polarization angle $\vartheta\in\left[0,\pi\right]$. With a solid radial line, we denote the angle corresponding to the maximum transition probability, measured to be $\vartheta\simeq 2.79\,$rad. 
 {\bf c)} Polar view of the transition probability associated with the excitation of the triton to the first excited state (VQD) extended for angles $\vartheta\in\left[0^\circ,360^\circ\right]$.
 With a solid radial line, we denote the angle corresponding to the maximum transition probability, measured to be $\vartheta\simeq -20^\circ$. 
 {\bf d)} The 3-dimensional plot of the same transition probability, obtained by adding an angle on the $xy$-plane and repeating the simulation. One can retrieve the two-dimensional plot by considering the intersection with the plane in $y=0$ represented in light blue in panel (d). For further details on the three-dimensional extension, refer to Additional Section A.}
    \label{fig:res_lcu}
\end{figure*}

Figure~\ref{fig:res_lcu}(a) illustrates the behavior of the success probability, while Figure~\ref{fig:res_lcu}(b) shows the corresponding transition probability. The LCU scheme consistently achieves a success probability above $30\%$. Such success probability has to be taken into consideration when running the algorithm on real quantum hardware. Since the number of shots for a single $\vartheta$ value is fixed, a lower success probability results in fewer samples for evaluating the corresponding transition probability. Figure~\ref{fig:res_lcu}(c) reports the results in light blue in polar coordinates. For the sake of completeness, we show that the behaviour is qualitatively different with respect to that of deuteron reported in Ref.~\cite{Roggero20C} (in grey). While both exhibit similar transition probability behavior, tritium peaks at $2.79\,$rad instead of $0\,$rad of the deuteron.\\
The simulation can be generalized to three-dimensional space by adding a polarization angle in the $xy$-plane without significant computational overhead. For a more detailed description of such generalization, refer to Equation~SE8 in the Additional Section A. Figure~\ref{fig:res_lcu}(d) shows such three-dimensional extension, with the magnitude representing the transition probability.

\section{Conclusions}
We processed the simulation of a nuclear transition involving totally antisymmetrized spin-isospin states of triton, formulated in terms of second quantization. The state of the nucleons is encoded by two qubits each on a gate model quantum computer.  We determined the ground state of the approximated triton Hamiltonian by exploiting the variational quantum eigensolver (VQE), which returned the energy expectation value with a relative error of $1\%$. In order to obtain the first excited state, two variations of the VQE have been deployed and compared, namely the Variational Quantum Deflation (VQD)~\cite{higgott19} and the VQE with automatically-adjusted constraints (VQE/AC)~\cite{Gocho23}, respectively. The two quantum states thus found have an overlap of $98\%$ with each other, suggesting the equal validity of the two methods. The relative errors on the corresponding eigenvalue are $9\%$ for the VQD and $8\%$ for the VQE/AC algorithm.
Finally, we simulated the transitions from the VQE/AC approximated excited state into the ground state with the LCU method. The simulation of the transition shows a success probability in the range $[0.3,0.9]$.
\\
This work provides a first step into a fully embedded quantum simulation, from state preparation to transition probability. Spatial resolution could also be embedded in the present framework by introducing additional qubits for each lattice site~\cite{Roggero20D}. 
Besides accounting for the details of the nuclear force entirely, similar simulations would also reach larger systems with similar computing effort as lighter systems on quantum hardware. These are not yet feasible due to resource constraints, but current research and proof-of-concept studies, such as those discussed above, provide a robust groundwork for progressing in this challenge.
%

\newpage
\section*{Supporting Information}
Supporting Information is available from the Wiley Online Library or from the author.
\section*{Acknowledgements}
E.P. acknowledges the project CQES of the Italian Space Agency (ASI). E.P. and L.N. thanks ENI S.p.A., for having partially supported this work.
\section*{Data availability}
The data that support the findings of this study are available from the corresponding author upon reasonable request.

\section*{Code availability}
The code and the algorithm used in this study are available from the corresponding author upon reasonable request.

\section*{Author Contributions}
L.N. developed the simulation and implemented the algorithms, C.B. elaborated on the physical interpretation of the quantum simulation in the realm of nuclear physics, E.P. conceived and coordinated this research. All the Authors contributed to discuss the results and to the writing of the manuscript.

\section*{Competing Interests statement}
The authors declare that there are no competing interests
\medskip

%
\bibliographystyle{MSP}



\begin{sidewaystable}[htbp]
    \centering
    \ra{1.2}
    \begin{tabular}{lccccccccccccccccc}\toprule
        &\phantom{a} & $\bm{\theta_{1}}$ & $\bm{\theta_{2}}$ & $\bm{\theta_{3}}$ & $\bm{\theta_{4}}$ & $\bm{\theta_{5}}$ & $\bm{\theta_{6}}$ & $\bm{\theta_{7}}$ & $\bm{\theta_{8}}$ & $\bm{\theta_{9}}$ & $\bm{\theta_{10}}$ & $\bm{\theta_{11}}$ & $\bm{\theta_{12}}$ & $\bm{\theta_{13}}$ & $\bm{\theta_{14}}$ & $\bm{\theta_{15}}$& $\bm{\theta_{16}}$ \\
        \cmidrule{3-18}
        \textbf{VQE}     & &3.844 & -0.681 & 6.510 & 3.526 & -4.452 & 7.411 & 4.764 & 5.181&-5.026 & 0.444 & -1.456 & 5.666 & 2.047 & 3.881 & -0.937 & -3.530\\
        \textbf{VQD}     & &4.891 & 1.247  & -4.682& -2.074& -3.245 & -1.542& 3.301& -5.139&-3.043 & 3.870 &4.424& 2.058 &-5.118&-1.609&4.680&5.944\\
        \textbf{VQE/AC}  & &3.206&-1.154&-1.938&-1.819&-2.934&1.126&4.010&-4.399&-2.199&4.315&4.011&0.284&-3.206&0.114&4.700&6.185\\
        \bottomrule
    \end{tabular}
    \caption{Optimal parameters obtained by COBYLA. With the variational quantum eigensolver (VQE)
    we find the parameters that let us prepare the ground state, whereas the variational quantum deflation (VQD) and the variational quantum eigensolver with automatically-adjusted constraint (VQE/AC) give the parameters that let us prepare the first excited state.}
    \label{tab:optimal_pars}
\end{sidewaystable}


%
\end{document}